\documentclass{pi2005}          

\usepackage{epsfig}

\usepackage{amsmath}
\usepackage{amssymb}
\usepackage{cite}

\slacs{.4ex}

\newcommand{\arctanh}{{\rm arctanh}\,}
\newcommand{\arcsinh}{{\rm arcsinh}\,}

\newcommand{\eps}{\varepsilon}
\newcommand{\al}{\alpha}
\renewcommand{\be}{\beta}
\newcommand{\ga}{\gamma}
\newcommand{\de}{\delta}
\newcommand{\om}{\omega}
\newcommand{\si}{\sigma}
\newcommand{\Ga}{\Gamma}
\newcommand{\De}{\Delta}

\DeclareMathOperator{\extdm}{d}
\newcommand{\extd}{\extdm \!}

\begin{document}

\title{Logarithmic corrections to the entropy of the exact string black hole}  
\authori{Daniel Grumiller\footnote{This research has been supported partially by
the project on scientific cooperation between the Austrian Academy
of Sciences and the National Academy of Sciences of Ukraine
No.~01/04 ``Quantum Gravity, Cosmology and Categorification'', as
well as by an Erwin-Schr\"odinger fellowship of the Austrian
Science Foundation (FWF), project J-2330-N08.}}
\addressi{Institute for Theoretical Physics, University of Leipzig,\\
Augustusplatz 10--11, D-04109, Germany}
\authorii{}     \addressii{}
\authoriii{}    \addressiii{}
\authoriv{}     \addressiv{}
\authorv{}      \addressv{}
\authorvi{}     \addressvi{}
%
\headauthor{Daniel Grumiller}            
\headtitle{Logarithmic corrections to the entropy of the exact string black hole}             
\lastevenhead{Daniel Grumiller: Logarithmic corrections to the entropy of the exact string black hole} 
\pacs{???}     

\maketitle \abstract{Exploiting a recently constructed target
space action for the exact string black hole, logarithmic
corrections to the leading order entropy are studied. There are
contributions from thermal fluctuations and from corrections due
to $\alpha^\prime>0$ which for the microcanonical entropy appear
with different signs and therefore may cancel each other,
depending on the overall factor in front of the action. For the
canonical entropy no such cancellation occurs. Remarks are made
regarding the applicability of the approach and concerning the
microstates. As a byproduct a formula for logarithmic entropy
corrections in generic 2D dilaton gravity is derived.}

\section{Entropy of black holes}

Entropy $S$ is a remarkable physical quantity and black holes with
a horizon of area $A$ are remarkable (astro)physical objects. A
full microscopical understanding of their combination, that is,
the entropy of black holes (BHs), is often considered as an
important mile stone on the road to quantum gravity. The famous
Bekenstein--Hawking relation $S_{\mathrm{BH}}=A/4$ receives (to
leading order logarithmic) corrections,
$S=S_{\mathrm{BH}}+\al\ln{S_{\mathrm{BH}}}+\mathcal{O}(1)$ (with
$\al=\mathrm{const.}$) from quantum gravity effects, but also from
thermal fluctuations around the equilibrium. For a review on BH
thermodynamics which discusses also logarithmic corrections to
entropy cf.~e.g.~\cite{Page:2004xp} and references therein. See \cite{reviews} for further reviews.

The purpose of this paper is to discuss in detail the entropy of
the exact string BH (ESBH) \cite{Dijkgraaf:1992ba} and to
calculate logarithmic corrections from thermal fluctuations as
well as from subleading effects of the string-coupling
$\al^\prime$. This is possible exploiting a target space action
for the ESBH constructed recently \cite{Grumiller:2005sq}. The
limitation of the current approach as well as the possibility of
the cancellation of both contributions is discussed.

Probably because the corresponding calculations often are
deceptively simple, there exists a large amount of literature
devoted to the study of logarithmic corrections to entropy of
various BHs, cf.~\cite{Frolov:1996hd, Kaul:2000kf,
Das:2001ic,Setare:2003pr} (and references therein) for a somewhat
randomly selected list of references. So the question arises
whether it is useful and interesting to add yet another example.
As to the usefulness, I attempt to clarify \textit{en passant}
some slightly confusing statements found in the literature
regarding the applicability of calculations concerning thermal
fluctuations, which may be of pedagogical value. Regarding the
question of interest, the BH under consideration is the ESBH, a
nonperturbative solution of 2D string theory, valid to all orders
in $\al^\prime$. Until recently not even the leading order entropy
could be calculated because no suitable target space action was
known, but now that it is available one can easily obtain not only
the leading order entropy but also logarithmic corrections to it.
In particular, the intricate interplay between $\al^\prime$
corrections and thermal fluctuations will turn out to be of
physical relevance.

As a byproduct a result for logarithmic corrections to entropy of
generic 2D dilaton BHs is presented.

\section{Logarithmic entropy corrections in generic 2D dilaton gravity}

\subsection{General features of thermal fluctuations}

A brief collection of basic thermodynamical quantities with an
explanation of our notation is contained in appendix \ref{app:A}.
The calculation of fluctuations below follows loosely
\cite{Das:2001ic}, but differs in certain aspects. For a given
mass $M$ sufficiently close to the equilibrium value $M_0$ one may
expand the microcanonical entropy $S_{\rm mc}(M)=S_{\rm
mc}(M_0)+(M-M_0)S'_{\rm mc}+\frac12\,(M-M_0)^2 S''_{\rm mc}+\dots$
perturbatively, where prime denotes differentiation with respect
to mass and setting $M=M_0$ afterwards. Physically the BH mass is
not fixed anymore but rather allowed to fluctuate around $M_0$, as
expected for a BH coupled to a heat bath, e.g.~provided by its own
Hawking radiation. While the microcanonical entropy is a function
of (fixed) mass, the canonical one depends on temperature and
therefore is more suitable for our purposes. Temperature to
leading order is given by $1/T=S'_{\rm mc}$, which makes the
integral in \eqref{eq:entropy11} a well-defined Gaussian one
provided $S''_{\rm mc}<0$. Neglecting contributions of order of
unity yields $S_{\rm c}=S_{\rm mc}-\frac12\,\ln{(-S''_{\rm mc})}$.
By virtue of \eqref{eq:entropy12} the relation $-S''_{\rm
mc}=1/(C_{\rm mc}T^2)$ holds. In the present work it will be
assumed that in equilibrium microcanonical and canonical entropies
coincide, $S_{\rm c}^{\rm eq}=S_{\rm mc}(M_0)=:S$, and also the
respective specific heats are equivalent, $C_{\rm mc}=C_V$. The
canonical entropy which includes thermal fluctuations due to
slight variations in the BH mass according to the previous
discussion reads
\begin{equation}
  \label{eq:entropy6}
  S_{\rm c} = S + \sfrac12\,\ln{\left(C_V T^2\right)} + \dots\,,
\end{equation}
provided that the specific heat is positive, $C_V>0$. The omitted
terms are of order of unity. In the context of BH thermodynamics
Eq.~\eqref{eq:entropy6} is of interest because it includes
logarithmic corrections to the area law.\footnote{Sometimes in the
literature instead of \eqref{eq:entropy6} a corresponding formula
for corrections of the microcanonical entropy is presented, where
the sign of the logarithmic term is reversed. However, since mass
is allowed to fluctuate such results are difficult to interpret.}
Note that in the range of applicability of Eq.~\eqref{eq:entropy6}
thermal fluctuations lead to an increase of canonical entropy,
$S_c>S$, as might have been anticipated on general grounds.

The approximation \eqref{eq:entropy6} is valid only if thermal
fluctuations are much larger than quantum fluctuations. The
simplest necessary bound\footnote{In the literature sometimes the
``Landau--Lifshitz bound'' \cite{lali} $T\gg 1/\tau$ is invoked,
where $\tau$ is the relaxation time. However, since the latter
typically is of order of Planck time this condition requires a
temperature much higher than Planck temperature, which is not
fulfilled for any quasi-classical BH solution of interest. The
derivation of this bound invokes the uncertainty relation for a
quasi-classical system, $\De_Q E\De_Q x>x/\tau$. But in the
present case $x$ is the energy which trivially commutes with the
Hamiltonian. Thus, the assumptions required to derive the
Landau--Lifshitz bound do not hold. Note also that the bound
\eqref{eq:entropy10} is stricter than the one which guarantees an
adequate description of BHs by means of thermodynamics
\cite{Preskill:1991tb}, $C_V\gg 1$, as long as the Hawking
temperature is below Planck temperature.} one can provide is given
by
\begin{equation}
  \label{eq:entropy10}
  \De_{T} M =\left<M^2\right>-\left<M\right>^2=C_V T^2 \gg \De_{Q} M \geq 1\,,
\end{equation}
where $\De_{T}$ and $\De_Q$ denote thermal and quantum
fluctuations, respectively. Note that condition
\eqref{eq:entropy10} assures that the logarithm in
\eqref{eq:entropy6} is positive. The assumption $\De_{Q} M\geq 1$
is plausible as the minimal quantum fluctuations of the BH mass
are likely to be of order of the Planck mass.  For certain systems
it is conceivable that $\De_Q M$ actually is much higher. If, for
instance, quantum fluctuations yield logarithmic corrections to
the entropy then it does not appear to be meaningful to apply
\eqref{eq:entropy6} in order to study thermal fluctuations on top
of them. Therefore, one ought to be very careful regarding
cancellations of logarithmic corrections. Related caveats have
been expressed in \cite{Page:2004xp}.

\subsection{Entropy in 2D dilaton gravity}

\paragraph{2D dilaton gravity} Generic dilaton gravity in 2D comprises a large
class of models, some of which are toy models invented for the
purpose of studying BH evaporation and conceptual issues regarding
quantum gravity \cite{Callan:1992rs,Russo:1992ht}, while others
stem from certain limits of higher dimensional theories, e.g.~the
Schwarzschild BH in spherically reduced Einstein gravity, toroidal
reduction \cite{Achucarro:1993fd} of the BTZ BH
\cite{Banados:1992wn} or Kaluza--Klein reduction of the
(super)gravity Chern--Simons term \cite{Guralnik:2003we}. Also 2D
type 0A/0B string theory yields a low energy effective action
which belongs to this set of theories,
cf.~e.g.~\cite{Takayanagi:2003sm,Davis:2004xi}. For the purpose of
this paper it is sufficient to recall four facts (for a
comprehensive review and references cf.~\cite{Grumiller:2002nm}):
\begin{enumerate}
\item Generic 2D dilaton gravity,\footnote{$R$ is the Ricci scalar,
$g$ the determinant of the metric (with Lorentzian signature), and
$\nabla$ the torsion free metric compatible derivative. The
reformulation of \eqref{eq:entropy15} as a first order theory
(related to a specific Poisson-$\si$ model) \cite{Schaller:1994es}
is very convenient already classically \cite{Klosch:1996fi} and
crucial at the quantum level, see appendix \ref{app:B} for some
references. The second order version \eqref{eq:entropy15} has been
introduced for instance in Refs.~\cite{Russo:1992yg}.
\label{fn:1}}
\begin{equation}
  \label{eq:entropy15}
  L=\frac{1}{4\pi G}\int_{\mathcal{M}_2}\extd^2x\sqrt{-g}
  \Bigl(\Phi R + U(\Phi)(\nabla \Phi)^2-2V(\Phi)\Bigr)
\end{equation}
involves two free functions $U$, $V$ depending on the dilaton
field $\Phi$ which define the model, but for simple
thermodynamical considerations only a particular (conformally
invariant) combination thereof, denoted by
\begin{equation}
  \label{eq:entropy14}
  w(\Phi):=\int^\Phi V(y)\E^{\int^y U(z)\extd z}\extd y\,,
\end{equation}
is of relevance. The multiplicative and additive ambiguities
implicit in the indefinite integrals my be absorbed by a
respective rescaling and shift of the mass definition. The
coupling constant $G$ is dimensionless and irrelevant in the
absence of matter; nevertheless, we will keep it as mass and
entropy both scale with $1/G$.
\item Entropy is proportional to the dilaton evaluated at the
(Killing) horizon, which must be a solution of $w(\Phi)+M=0$,
where $M$ is essentially the BH mass. This statement will be
recalled in detail below by various methods.
\item Temperature is proportional to the first derivative of
$w(\Phi)$ evaluated at the horizon (this coincides with surface gravity).
\item Specific heat is proportional to temperature over the
second derivative of $w(\Phi)$ evaluated at the horizon. The
latter will be denoted by $r=w''$ since it corresponds to
curvature on the horizon in a conformal frame where $U=0$. The
specific heat is positive if the signs of the first and second
derivative of $w(\Phi)$ coincide at the horizon.
\end{enumerate}

\paragraph{Entropy from Wald's Noether charge technique}

Wald's method \cite{Wald:1993nt} to derive the BH entropy has been
applied numerously. Because all classical solutions of
\eqref{eq:entropy15} have at least one Killing vector this method
is very suitable here. The basic idea is inspired by standard
quantum field theory methods: An infinitesimal diffeomorphism
generated by some vector field $\xi$ yields for the variation of
the Lagrangian 2-form $L$ a term proportional to the equations of
motion plus an exact form denoted by $\extd\Theta$, where $\Theta$
is a 1-form. The 1-form $j=\Theta-\xi\cdot L$, where $\cdot$ means
contraction (with the first index), on-shell is not only closed
but even exact, $j=\extd Q$, where $Q$ is the Noether charge. If
$\xi$ coincides with the Killing vector associated with
stationarity (normalized to unity at infinity for asymptotically
flat spacetimes) and $Q$ is evaluated at the Killing horizon, then
the quantity $Q/T$ is the entropy of the BH, where $T$ is the
Hawking temperature as derived from surface gravity. For 2D
dilaton gravity, where $\xi^\mu\propto
\eps^{\mu\nu}\nabla_\nu\Phi$ and
$Q\propto\eps^{\mu\nu}(2\xi_\mu\nabla_\nu\Phi+\Phi\nabla_\mu\xi_\nu)$,
this method has been applied in \cite{Gegenberg:1995pv},
establishing the important result
\begin{equation}
  \label{eq:entropy21.5}
  S=\frac{1}{G}\,\Phi\bigg|_{w(\Phi)+M=0}=\frac{1}{G}\,\Phi_h\,.
\end{equation}
The famous Bekenstein--Hawking relation is encoded in
\eqref{eq:entropy21.5} because the dilaton field evaluated at the
horizon is the 2D equivalent of an ``area'' (for spherically
reduced gravity one can take this literally as $\Phi$ becomes
proportional to the surface area; see, however, below). The same
result can be obtained by virtue of the first law
\eqref{eq:entropy2}: $\extd M=-w'(\Phi)\extd\Phi=T\extd S\propto
-w'(\Phi)\extd S$, which implies $(S-S_0)\propto\Phi$ (all
$\Phi$-dependent quantities have to be evaluated at the horizon).
If one takes into account the appropriate proportionality factor
and sets the ``zero point entropy'' $S_0=0$ then
\eqref{eq:entropy21.5} is recovered. Thus, in 2D dilaton gravity
there are no corrections to the BH entropy from the Noether charge
technique (but there are examples in the literature where such
corrections arise, cf.~e.g.~\cite{LopesCardoso:1998wt}).

\paragraph{CFT counting of microstates} There exist various ways
to count the microstates by appealing to the Cardy formula and to
recover the result \eqref{eq:entropy21.5}. However, the true
nature of these microstates remains unknown in this approach,
which is a challenging open problem. Here are references, some of
which contain mutually contradicting results:
\cite{Strominger:1998yg}.

\paragraph{A poor man's counting of microstates} This is a speculative
paragraph, but I think it contains a grain of ``truth''. It could
be that the identification of the appropriate microstates in 2D
dilaton gravity has been unsuccessful so far because the problem
has been misinterpreted. What I mean is the following: in $D$
dimensions the Bekenstein--Hawking formula for entropy reads
$$
S=\frac{A}{4G}\,,
$$
supposing that there is a coupling constant $1/G$ in front of the
geometric action (times an appropriate numerical factor); even
though one might scale $G$ to unity it is kept for illustration.
Concerning the 2D dilaton gravity result \eqref{eq:entropy21.5}
there are two possibilities: $\Phi_h$ determines the area of the
horizon or $\Phi_h$ rescales the inverse coupling. From the
spherically reduced point of view the first interpretation is
apparent (as pointed out above $\Phi$ is essentially the surface
area), but for intrinsically 2D dilaton gravity only the second
possibility is applicable, as the area of a ``one-sphere'', $A=2$
(obtained by analytic continuation from
$A=2r^{D-1}\pi^{D/2}/\Ga(D/2)$ to $D=1$), is just a constant of
order of unity. This implies that the dilaton field is not related
to the area but rather to the effective gravitational coupling.
This concurs with the observation \cite{bgkv} that the dilaton
multiplies all boundary terms and thus the number of intrinsically
2D microstates has to be only of order of unity to produce the
correct formula for entropy. Note that the equation
\begin{equation}
G^{\rm eff} = \frac{G}{\Phi_h} = G\E^{2\phi_h}
\end{equation}
is well-known from string theory, where $\phi_h$ is the dilaton
field usually employed in string theory, evaluated at the horizon.

A macroscopic way to get 1 bit of microstates may be related to
the fact that globally a 1-horizon spacetime consists of two
copies of basic Eddington--Finkelstein patches, i.e., contains two
asymptotic regions.\footnote{This is a global consideration, but
as locally our theory is trivial global considerations are needed,
unless one couples matter to the system (which would be more
physical anyhow, but also more complicated).} So entropy arises
not for any surface but just for horizons. Up to numerical factors
of $\ln{2}$ and $\pi$ this provides the correct entropy if one
takes into account the effective coupling constant:
\begin{equation}
S=\frac{\mathcal{O}(1)}{G^{\mathrm{eff}}}=\frac{\mathcal{O}(1)}{G}\,\Phi_h\,.
\end{equation}
However, I don't know a way to actually derive the correct
numerical prefactor. It were marvellous if the ideas expressed
above could be put on firmer grounds.

\subsection{Log corrections in generic 2D dilaton gravity}

\paragraph{Thermal fluctuations}
Putting the previous observations together, Eq.~\eqref{eq:entropy6} yields
\begin{equation}
  \label{eq:entropy16}
  S_{\rm c} = \frac{1}{G}\Phi+\frac32\ln{T}-\frac12\ln|r|+\dots\,,
\end{equation}
recalling that $T\propto |w'|$ and $r=w''$ evaluated at the
horizon. A large and interesting class of models is given by
$w=-\Phi^\alpha$, including the $D$-dimensional Schwarzschild BH
($\al=(D-3)/(D-2)$), the Witten BH ($\al=1$)
\cite{Mandal:1991tz,Callan:1992rs} and the Jackiw--Teitelboim
model ($\al=2$) \cite{Teitelboim:1983ux}. Insertion into
\eqref{eq:entropy16} gives
\begin{equation}
  \label{eq:entropy17}
  S_{\rm c}=S+\left(\al-\sfrac12\right)\ln{S} + \dots
\end{equation}
with $S=M^{1/\al}/G\propto T^{\al/(\al-1)}$. Amusingly, for the 4D
Schwarzschild BH the corrections vanish, but since $C_V<0$ in that
case (and even if the sign of $C_V$ was reversed the bound
\eqref{eq:entropy10} is violated), Eq.~\eqref{eq:entropy6} is not
applicable. In fact, positivity of the specific heat requires
$\al>1$, so that the coefficient in front of the logarithmic term
in \eqref{eq:entropy17} is always positive and bigger than
$\frac12$. The condition \eqref{eq:entropy10} demands
$M^{2-1/\al}\gg(\al-1)/\al^2$. So in general the mass of the BH
has to be large, as expected intuitively.

\paragraph{Semiclassical corrections} If matter is coupled to
the geometric action \eqref{eq:entropy15} one may apply 1-loop
results to obtain logarithmic corrections from quantum
fluctuations of matter on a given background \cite{Frolov:1996hd}
(see also \cite{Fiola:1994ir}). This approach is interesting on
its own but it will not be considered here.

\section{Logarithmic corrections for the ESBH}

\subsection{A mini review on the ESBH}

\paragraph{Geometry of the ESBH}  The ESBH geometry was discovered by
Dijkgraaf, Verlinde and Verlinde more than a decade ago
\cite{Dijkgraaf:1992ba}.\footnote{At the perturbative level
actions approximating the ESBH are known: to lowest order in
$\alpha^\prime$ an action emerges the classical solutions of which
describe the Witten BH \cite{Mandal:1991tz}, which in turn
inspired the CGHS model \cite{Callan:1992rs}, a 2D dilaton gravity
model with scalar matter that has been used as a toy model for BH
evaporation. Pushing perturbative considerations further Tseytlin
was able to show that up to 3 loops the ESBH is consistent with
sigma model conformal invariance \cite{Tseytlin:1991ht}; in the
supersymmetric case this holds even up to 4 loops
\cite{Jack:1993mk}. In the strong coupling regime the ESBH
asymptotes to the Jackiw--Teitelboim model
\cite{Teitelboim:1983ux}. The exact conformal field theory (CFT)
methods used in \cite{Dijkgraaf:1992ba}, based upon the
$\mathrm{SL}(2,\mathbb{R})/\mathrm{U}(1)$ gauged
Wess--Zumino--Witten model, imply the dependence of the ESBH
solutions on the level $k$. A different (somewhat more direct)
derivation leading to the same results for dilaton and metric was
presented in \cite{Tseytlin:1992ri} (see also \cite{Bars:1993zf}).
For a comprehensive history and more references
ref.~\cite{Becker:1994vd} may be consulted.} In the notation of
\cite{Kazakov:2001pj} for Euclidean signature the line element of
the ESBH is given by
\begin{equation}
\extd s^2=f^2(x)\extd\tau^2+\extd x^2\,, \label{eq:dvv1}
\end{equation}
with
\begin{equation}
f(x)=\frac{\tanh(bx)}{\sqrt{1-p\tanh^2(bx)}}\,. \label{eq:dvv2}
\end{equation}
Physical scales are adjusted by the parameter $b\in\mathbb{R}^+$
which has dimension of inverse length. The corresponding
expression for the dilaton,
\begin{equation}
\phi=\phi_0-\ln\cosh(bx)-\sfrac14\,\ln\left(1-p\tanh^2(bx)\right),
\label{eq:dvv3}
\end{equation}
contains an integration constant $\phi_0$. Additionally, there are
the following relations between constants, string-coupling
$\al^\prime$, level $k$ and dimension $D$ of string target space:
\begin{equation}
\alpha^\prime b^2=\frac{1}{k-2}\,,\qquad
p:=\frac{2}{k}=\frac{2\alpha^\prime b^2}{1+2\alpha^\prime
b^2}\,,\qquad D-26+6\alpha^\prime b^2=0\,.
\label{eq:dvv4}
\end{equation}
For $D=2$ one obtains $p=\frac89$, but like in the original work
\cite{Dijkgraaf:1992ba} we will treat general values of
$p\in(0;1)$ and consider the limits $p\to 0$ and $p\to 1$
separately: for $p=0$ one recovers the Witten BH geometry; for
$p=1$ the Jackiw--Teitelboim model is obtained. Both limits
exhibit singular features: for all $p\in(0;1)$ the solution is
regular globally, asymptotically flat and exactly one Killing
horizon exists. However, for $p=0$ a curvature singularity
(screened by a horizon) appears and for $p=1$ space--time fails to
be asymptotically flat. In the present work exclusively the
Minkowskian version of (\ref{eq:dvv1})
\begin{equation}
\extd s^2=f^2(x)\extd\tau^2-\extd x^2\,, \label{eq:dvv5}
\end{equation}
will be needed. The maximally extended space--time of this
geometry has been studied by Perry and Teo \cite{Perry:1993ry} and
by Yi \cite{Yi:1993gh}. Winding/momentum mode duality implies the
existence of a dual solution, the Exact String Naked Singularity
(ESNS), which can be acquired most easily by replacing $bx\to
bx+\I\pi/2$, entailing in all formulas above the substitutions
\begin{equation}
\sinh\to \I\cosh\,,\qquad \cosh\to \I\sinh\,. \label{dvv:dual}
\end{equation}

\begin{figure}
\centering \epsfig{file=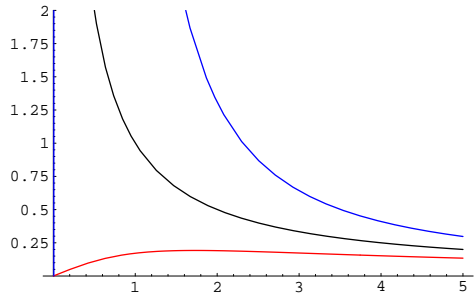,width=.5\linewidth}
\vskip-5mm \caption{The potential $U$ as function of the auxiliary
dilaton $\ga$.} \label{fig1}
\end{figure}

\paragraph{A target space action for the ESBH} It took surprisingly
long until a suitable target space action had been constructed
\cite{Grumiller:2005sq}. Typically this means that either the
problem is very difficult or not very interesting. The problem of
finding a target space action and its importance for entropy has
been addressed already in early publications,
cf.~e.g.~\cite{Gibbons:1992rh}, as well as in recent ones
\cite{Davis:2004xi}. This may be taken as an indication for the
first option. In rare cases, however, there is a third one, namely
that the solution is actually simple once appropriate tools are
employed. Indeed, after it had been realized that the nogo result
of \cite{Grumiller:2002md} may be circumvented without introducing
superfluous physical degrees of freedom simply by adding an
abelian $BF$-term, a straightforward reverse-engineering procedure
allowed to construct uniquely a target space action of the form
\eqref{eq:entropy15}, supplemented by aforementioned $BF$-term.
{\em Per constructionem} it reproduces as classical solutions
precisely Eqs.~\eqref{eq:dvv2}--\eqref{eq:dvv5} not only locally
but globally. Incidentally also the first order formulation has
been a pivotal technical prerequisite. The corresponding first
order Maxwell-dilaton gravity action (which for $BF=0$ is
classically equivalent to \eqref{eq:entropy15} with the same
functions $U$ and $V$)\footnote{The notation is explained in more
detail in \cite{Grumiller:2005sq}. For sake of self-containment
here is a brief summary: the 2-forms $T^\pm=(\extd\pm\om)\wedge
e^\pm$, $R=\extd\om$ and $F=\extd A$ are torsion, curvature, and
abelian field strength, respectively and depend on the gauge field
1-forms $e^\pm$ (``Zweibein''), $\omega$ (``spin connection'') and
$A$ (``Maxwell field''). The scalar fields $X_\pm$, $\Phi$ and $B$
are Lagrange multipliers for these 2-forms and appear also in the
potential, the last term in \eqref{eq:solutionofESBH}, which is
multiplied by the volume 2-form $\epsilon=e^+\wedge e^-$. The
$\pm$ indices refer to lightcone gauge for the anholonomic frame,
i.e., the flat metric raising and lowering such indices reads
$\eta_{\pm\mp}=1$, $\eta_{\pm\pm}=0$. The overall normalization of
\eqref{eq:solutionofESBH} differs from \cite{Grumiller:2005sq} by
$1/(2\pi G)$ for sake of consistency with Eq.~\eqref{eq:entropy15}.}
\begin{equation}
  \label{eq:solutionofESBH}
L_{\mathrm{ESBH}}=\frac{1}{2\pi G}\int_{{\mathcal M}_2}
\Bigl[X_aT^a+\Phi R+BF+\epsilon
\bigl(X^+X^-U(\Phi)+V(\Phi)\bigr)\Bigr]\,,
\end{equation}
with potentials $U$, $V$ to be defined below, describes the ESBH
as well as the ESNS, i.e., on-shell the metric
$g_{\mu\nu}=\eta_{ab}\,e^a_\mu\, e^b_\nu$ and the dilaton $\phi$
are given by \eqref{eq:dvv2}--\eqref{dvv:dual}. Regarding the
latter, the relation
\begin{equation}
  \label{eq:solutionofESBH1}
  (\Phi-\gamma)^2 = \arcsinh^2{\gamma}
\end{equation}
in conjunction with the definition
\begin{equation}
  \label{eq:solutionofESBH2}
  \gamma:=\frac{\E^{2\phi}}{B}
\end{equation}
may be used to express the auxiliary dilaton field $\Phi$ in terms
of the ``true'' dilaton field $\phi$ and the auxiliary field $B$.
The two branches of the square root function correspond to the
ESBH (main branch) and the ESNS (second branch), respectively.
Henceforth the notation (to be distinguished from lightcone
indices!)
\begin{equation}
  \label{eq:solutionofESBH2.5}
  \Phi_\pm = \gamma \pm \arcsinh{\gamma}
\end{equation}
will be employed, where $+$ refers to the ESBH and $-$ to the
ESNS. The potentials read
\begin{equation}
  \label{eq:solutionofESBH3}
  V=-2b^2\gamma\,,\qquad U_\pm= -\frac{1}{\gamma N_\pm(\gamma)}\,,
\end{equation}
with an irrelevant scale parameter $b\in\mathbb{R}^+$ and
\begin{equation}
  \label{eq:solutionofESBH4}
  N_\pm(\gamma)=1+\frac{2}{\gamma}\left(\frac{1}{\gamma}\pm\sqrt{1+\frac{1}{\gamma^2}}\right).
\end{equation}
Note that $N_+N_-=1$. This completes the definition of all terms
appearing in the action \eqref{eq:solutionofESBH}. In figure
\ref{fig1} the potential $U$ is plotted as function of the
auxiliary dilaton $\ga$. The lowest branch is associated with the
ESBH, the one on top with the ESNS and the one in the middle with
the Witten BH. The regularity of the ESBH is evident, as well as
the convegence of all three branches for $\ga\to\infty$. The
potential $V$ is linear and homogeneous in $\ga$ for all three
branches, see Eq.~\eqref{eq:solutionofESBH3}.

For evaluation of entropy the auxiliary dilaton $\Phi$ is the
relevant field because it multiplies the curvature scalar in the
action and therefore the (Gibbons--Hawking) boundary term contains
$\Phi$ evaluated at the boundary \cite{bgkv}. Then one may
trivially integrate out the abelian $BF$ term as it decouples from
all other fields.

\paragraph{Recollection of relevant quantities}
The (target space) action \eqref{eq:solutionofESBH} is needed in
order to be able to discuss entropy of the ESBH. The conformally
invariant combination of the dilaton potentials defined in
\eqref{eq:entropy14} reads\footnote{This may be obtained directly
from \eqref{eq:solutionofESBH3} by fixing conveniently the scaling
and shift ambiguity.}
\begin{equation}
  \label{eq:entropy18}
  w=-b\left(1+\sqrt{1+\ga^2}\right) \,,
\end{equation}
where $\ga$ is an auxiliary field which fulfils (see Eq.~\eqref{eq:solutionofESBH2.5})
\begin{equation}
  \label{eq:entropy19}
  \frac{\extd\ga}{\extd\Phi}=1+\frac{1}{\sqrt{1+\ga^2}}\,.
\end{equation}
The ADM mass is given by
\begin{equation}
  \label{eq:ADM}
  M_{\rm ADM} = \frac{b}{2\pi G}\,k = \frac{2b}{2\pi G}\,M \,,
\end{equation}
where $k$ is the level, $b$ an irrelevant scale parameter and the
dimensionless quantity $M$ used henceforth lies in the interval
$(1,\infty)$. Hawking temperature reads
\begin{equation}
  \label{eq:dvvnew20}
  T=\frac{b}{2\pi}\,\sqrt{1-\frac{1}{M}} \,.
\end{equation}
The microcanonical entropy in equilibrium takes the form
\begin{equation}
  \label{eq:intentropy}
  S=\frac{2}{G}\left(\sqrt{M(M-1)}+\arctanh{\sqrt{1-\frac1M}}\right).
\end{equation}
Finally, the specific heat will be needed:
\begin{equation}
  \label{eq:cs}
  C_V= \frac{8\pi}{b G}\,M^2\,T>0\,.
\end{equation}

\subsection{Thermal fluctuations vs.~\bmth{\al^\prime} corrections}

Application of \eqref{eq:entropy16} is possible only if $M\gg 1$.
Thus, in order to study thermal fluctuations it is sufficient to
expand  \eqref{eq:intentropy} around $M=\infty$. Therefore,
already the \textit{equilibrium} entropy acquires logarithmic
corrections which are $\al^\prime$ corrections:\footnote{The limit
$M\gg 1$ implies $k\gg 1$ and via \eqref{eq:dvv4} also
$\al^\prime\ll 1/b^2$. Thus, $\al^\prime$ has to be small.}
\begin{equation}
  \label{eq:entropy21}
  S=\frac{2}{G}\left(M+\frac12\ln{M}+\dots\right).
\end{equation}
The leading order term will be denoted by
$S^{\mathrm{WBH}}=2M/G=2/((1-(2\pi T/b)^2)G)$. Note that the
subleading term comes with a \textit{positive sign}. It is
emphasized that neither thermal nor quantum fluctuations have been
considered so far. If the latter are negligible, i.e., $M \gg 1$,
and additionally $\ln{M}\gg|\ln{G}|$ then Eq.~\eqref{eq:entropy16}
together with \eqref{eq:entropy21} establishes our main result,
\begin{equation}
  \label{eq:entropy20}
  S^{\mathrm{ESBH}}_{\mathrm{c}}=S^{\mathrm{WBH}}+\left(1+\frac{1}{G}\right)\ln S^{\mathrm{WBH}}\,.
\end{equation}
It should be noted that for corresponding corrections to the
microcanonical entropy, as mentioned below \eqref{eq:entropy6},
the term arising from thermal fluctuations reverses its sign,
i.e.,
\begin{equation}
  \label{eq:entropy28}
  S^{\mathrm{ESBH}}_{\rm mc}=S^{\mathrm{WBH}}-\left(1-\frac{1}{G}\right)\ln{S^{\mathrm{WBH}}}\,.
\end{equation}

\subsection{Discussion}

While for the canonical entropy both corrections \textit{increase}
the entropy, for the microcanonical entropy the following happens:
If $G$ is small [large] the $\al^\prime$ corrections [thermal
fluctuations] dominate and yield a \textit{positive}
[\textit{negative}] prefactor in front of the logarithmic term. If
$G=1$ both kind of corrections cancel each other and the
``improved'' microcanonical entropy of the exact string BH is
equal to $S_{\mathrm{WBH}}$, i.e., proportional to the mass. It is
stressed that this is a consistent cancellation of logarithmic
contributions in the sense that it is not resulting from thermal
fluctuations on top of quantum fluctuations, but rather from
thermal fluctuations on top of $\al^\prime$ corrections. However,
as already mentioned the use of the microcanonical entropy is
difficult to interpret physically if mass is allowed to fluctuate.
In any case, the appearance of the dimensionless coupling constant
$G$ in \eqref{eq:entropy20} due to $\al^\prime$ corrections seems
to be a remarkable new feature in the context of log-corrections
to BH entropy.

If $E-TS\neq 0$ there are two possibilities: either work terms are
present or energy contains a ``Casimir contribution'' violating
the Euler relation $E=TS$ \cite{Verlinde:2000wg}, i.e., a
non-extensive component which in $D$ dimensions scales as
$E_C(S/G)=E_C(S)/G^{1-2/D}$, as opposed to the extensive part
$E(S/G)=E(S)/G$. In 2D $E_C$ does not scale at all (or at most
logarithmically). If one defines\footnote{The choice
$E=M_{\mathrm{ADM}}$ is not natural because the ground state
solution $M\to 1$ has $M_{\mathrm{ADM}}=b/(\pi G)$. The definition
in the text yields vanishing $E$ for the ground state.}
$E=M_{\mathrm{ADM}}-b/(\pi G)$ one obtains in the absence of
fluctuations\footnote{As the fluctuations scale with $\ln{G}$ they
may be considered as a ``Casimir contribution'' to the energy.
This seems to be a peculiar feature of BHs in 2D only.}
\begin{equation}
  \label{eq:entropy29}
  E-TS=-\frac{b}{\pi G}\sqrt{1-\frac{1}{M}}\,\arctanh{\sqrt{1-\frac{1}{M}}}\approx -\frac{T}{G}\ln{S^{\mathrm{WBH}}}\,.
\end{equation}
Because the right hand side above scales extensively with $1/G$ it
seems adequate to infer the presence of work
terms.\footnote{Normally ``extensive'' refers to a scaling with
the volume. However, this is not applicable for an isolated
asymptotically flat BH, unless one considers e.g.~the combined
system BH plus Hawking radiation and puts the BH into a cavity of
finite size. Therefore, the only scale available is the coupling
constant $G$. Since both energy and leading order entropy scale
with $1/G$ while temperature does not scale, the attribute
``extensive'' is appropriate. On a sidenote, the Euler identity in
the presence of work terms reads $E-TS+A_iB_i=0$ where all $A_i$
are intensive (e.g.~pressure $p$) and $B_i$ extensive (e.g.~volume
$V$). However, the condition $\extd M=T\extd S$ has emerged in the
derivation of entropy -- thus, if work terms were present one
ought to reconsider also the derivation of entropy.
Cf.~e.g.~Ref.~\cite{Davis:2004xi} for a recent discussion of the
thermodynamics of 2D BHs in a cavity.} They vanish for $E\to 0$.
It could be quite interesting to unravel their physical meaning.
We will provide but the first step: To this end it is helpful to
observe that the right hand side in \eqref{eq:entropy29} is
nothing but (minus) temperature times the $\al^\prime$ correction
to the (micro-)canonical entropy. Thus, the $\al^\prime$
corrections are responsible for the violation of $E=TS$.

Finally, it is recalled that the result \eqref{eq:entropy16} may
be applied to any 2D dilaton gravity model and therefore could be
useful for future studies.

\bigskip
\noindent {\small I am grateful to D.~Vassilevich for discussion
and to S.~Moskaliuk for important administrative help. Some of the
results have been presented at a conference in Prague ``Path
Integrals. From Quantum Information to Cosmology'' in June 2005.
The relevant references concerning my talk ``Path integral
quantization of the ESBH'' are contained in appendix \ref{app:B}.}

\begin{appendix}

\section{Basic thermodynamical notions}\label{app:A}

Entropy may be defined as the maximum of the mean expected
information under given constraints, times an irrelevant
dimensionful factor, the Boltzmann constant, which will be set to
unity in the present work.\footnote{Also all other conversion
factors ($c,\hbar,\dots$) will be set to unity.} Although not
always a detailed microscopic understanding is available for a
given system, especially in BH physics, it is nevertheless useful
to adopt the information theoretic approach: For a given discrete
ensemble with probability distribution $w_\nu$ (where $\sum_\nu
w_\nu=1$, the sum extending over all possible configurations
$\nu=1,\dots,N$ and $w_\nu\in[0,1]$ being the probability to
measure a particular configuration $\nu$) the information for any
event $v$ is defined as $-\ln{w_\nu}$. Its mean value is
$-\sum_\nu w_\nu\ln{w_\nu}$. The canonical entropy in the context
of BH physics rests upon the physical constraint $\sum_\nu w_\nu
M_\nu=:\left<M\right>=M$:
\begin{equation}
  \label{eq:entropy1}
  S_{\rm c}=\max\left[-\sum_\nu w_\nu\ln{w_\nu}-\al\biggl(\sum_\nu w_\nu-1\biggr)-
  \be\biggl(\sum_\nu w_\nu M_\nu-M\biggr)\right].
\end{equation}
The quantity $\exp{(1+\al)}$ is called partition function and
usually is denoted by $Z$. The quantity $T=1/\beta$ is called
(Hawking) temperature of the BH. Ensuingly,
$S_{\mathrm{c}}=\ln{Z}+M/T$. From $\de w_\nu(\ln{w_\nu}+1+\al+\be
M_\nu)=0$ one deduces $w_\nu=\exp{(-M_\nu/T)}/Z$. The first law of
thermodynamics/BH mechanics,
\begin{equation}
  \label{eq:entropy2}
 \extd M = T \extd S_{\rm c}\,,
\end{equation}
is a simple consequence. Work terms are assumed to be absent,
which is sufficient for our purposes.

The microcanonical entropy, $S_{\rm mc}$, may be obtained as a
special case from the canonical entropy where only states with the
same value of $M_{\nu^\prime}$ are considered,
$S_{\mathrm{mc}}=\ln\rho$, where
$\rho=1/w_{\nu^\prime}=Z\exp{(M_{\nu^\prime}/T)}$ is the
corresponding number of microstates (which necessarily are equally
distributed). Apart from rare but interesting exceptions the
equilibrium values of canonical and microcanonical entropies
coincide.\footnote{For systems where this ceases to be true
cf.~e.g.~\cite{Touchette:2003} and references therein. See
footnote \ref{fn:9}.} Such an exception arises, for instance, if
the microcanonical specific heat,
\begin{equation}
  \label{eq:entropy12}
  C_{\rm mc}=-\frac{\left(S'_{\rm mc}\right)^2}{S''_{\rm mc}}\,,
\end{equation}
is negative, where prime denotes differentiation with respect to $M$.

In the continuum case sums are replaced by integrals and $\rho(M)$
becomes the density of states at a given mass $M$. The partition
function may be considered as the Laplace transformed density of
states,
\begin{equation}
   \label{eq:entropy11}
   Z(\beta)=\int_0^\infty\rho(M)\E^{-\beta M}\extd M=
   \int_0^\infty \E^{S_{\rm mc}(M)-\beta M}\extd M\,.
 \end{equation}
This formula allows to derive corrections to the partition
function from thermal fluctuations in the mass.

If there are two subsystems which may exchange energy (like a BH
and radiation) then the total entropy $S_{\rm tot}$ is the maximum
of the sum of the entropies $S_1$, $S_2$ of the subsystems under
the constraint that $E_1+E_2=E$, where $E$ is a constant, the
total energy, and $E_1$, $E_2$ are the energies of the subsystems.
Variation under such a constraint yields that the temperatures of
both subsystems have to be equal. However, equality of
temperatures only establishes an extremum of the total entropy.
For a maximum in addition the inequality
\begin{equation}
  \label{eq:entropy3}
  \frac{\partial^2 S_1}{\partial E_1^2} + \frac{\partial^2 S_2}{\partial E_2^2} < 0\,,
\end{equation}
has to be fulfilled. Defining the (canonical) specific heat as
\begin{equation}
  \label{eq:entropy4}
  C_V := -\beta^2\,\frac{\partial^2\ln{Z}}{\partial\beta^2}=
  T\,\frac{\partial S_{\rm c}}{\partial T}\,,
\end{equation}
one can see that \eqref{eq:entropy3} holds if the specific heat of
each subsystem, in particular of the BH, is positive.\footnote{A
well-known counter example is the Schwarzschild BH which has
negative specific heat, $C_{\rm mc}\propto -M^2$. This implies
that a Schwarzschild BH can only be in thermal equilibrium with
radiation if the volume filled by the latter is not too large,
$V<({\rm const.})M^5$, such that \eqref{eq:entropy3} is not
violated. \label{fn:9}}

\section{Path integral quantization of the ESBH}\label{app:B}

This appendix is much closer to the actual content of my talk at
the conference ``Path Integrals: From Quantum Information to
Cosmology'' in Prague, but as it does not contain essential new
considerations I shall be very brief and provide mainly a pointer
to the literature. For a first orientation and a more
comprehensive list of references the review article
Ref.~\cite{Grumiller:2002nm} may be consulted. Below the emphasis
will be on the ``Vienna School approach''. The basic feature
discriminating it from other approaches, like Dirac quantization
(cf.~e.g.~\cite{Cangemi:1992bj}), is the possibility to integrate
out geometry \textit{exactly} even in the presence of matter, thus
obtaining an effective theory with nonlocal and nonpolynomial
matter interactions. The latter is a suitable starting point for a
perturbative treatment where to each given order geometry may be
reconstructed self-consistently. The path integral formulation
appears to be the most adequate language to derive these
perturbative results.

Various aspects of the path integral quantization of generic
dilaton gravity (including different signatures and supergravity
extensions) with matter have been discussed in a series of
publications \cite{Haider:1994cw} which, in a sense, are built
upon the two basic papers of Kummer and Schwarz
\cite{Kummer:1992bg}. In particular, $S$-matrix calculations where
``Virtual Black Holes'' emerge as intermediate states have been
performed and discussed in
Refs.~\cite{Grumiller:2000ah,Grumiller:2001ea,Grumiller:2002dm,Grumiller:2004yq}.
Previous proceedings contributions may also be of use as a further
literature guide \cite{Grumiller:2001pt}. It is emphasized that
the first order formulation mentioned in footnote \ref{fn:1} has
been a crucial technical ingredient.

Applying this formalism to the ESBH is possible since a suitable
target space action has been constructed \cite{Grumiller:2005sq}.
If the geometric action \eqref{eq:entropy15} is supplemented by
scalar matter (e.g.~the tachyon)
 \begin{equation}
   \label{eq:entropy100}
   L^{(m)}=\int_{\mathcal{M}_2} \extd^2x\sqrt{-g}\Bigl(F(\Phi)(\nabla\phi)^2+f(\Phi,\phi)\Bigr)\,,
 \end{equation}
where $F$ is the coupling function and $\phi$ the scalar field,
exact solutions emerge only for very specific cases as the system
now exhibits one propagating physical degree of freedom. The
considerations of local self interactions encoded in the function
$f$ is straightforward and for simplicity will be omitted here.
Since the conference has been on Path Integrals it is appropriate
to provide the latter at least schematically\footnote{The quantity
$W$ is the generating functional for Green functions. The term
``ghosts'' denotes the whole ghost and gauge-fixing sector.
Gauge-fixing has been chosen such that Eddington--Finkelstein (or
Sachs--Bondi) gauge is produced for the line element. It should be
noted that the path integral \eqref{eq:pifull} involves positive
and negative values of the dilaton and both orientations. Further
details on the quantization procedure may be found in appendix E
of \cite{Grumiller:2001ea} and in Section 7 of
\cite{Grumiller:2002nm}. This paragraph essentially has been
copied from \cite{Grumiller:2004yq}.}
\begin{equation}
\label{eq:pifull}
\begin{array}{rcl}
W(\mathrm{sources})&=&\disty\int\mathcal{D}(\mathrm{geometry})\,\mathcal{D}(\mathrm{dilaton})\,
\mathcal{D}(\mathrm{ghosts})\,\mathcal{D}\phi\times\\[9pt]
&&\disty\times\exp\left[\I\int\extd^2x
\left(\mathcal{L}^{\mathrm{eff}}(\mathrm{geometry},\mathrm{dilaton},\mathrm{ghosts},\phi)+
\mathrm{sources}\right)\right].
\end{array}
\end{equation}
Path integration over all fields but matter is possible exactly,
{\em without} introducing a fixed background geometry. Thus, the
quantization procedure is non-perturbative and background
independent. However, there are ambiguities coming from
integration constants the fixing of which selects a certain
asymptotics of spacetime; two of them are trivial while the third
one essentially determines the ADM mass (whenever this notion
makes sense).\footnote{The issue of mass is slightly delicate in
gravity. For a clarifying discussion in 2D see
\cite{Liebl:1997ti}. One of the key ingredients is the existence
of a conserved quantity \cite{Banks:1991mk} which has a deeper
explanation in the context of first order gravity
\cite{Grosse:1992vc} and Poisson-$\si$ models
\cite{Schaller:1994es}. A recent mass definition extending the
range of applicability of \cite{Liebl:1997ti} may be found in
appendix A of \cite{Grumiller:2004wi}. The relations between this
conserved quantity, the ADM mass, the Bondi masses and the
Misner--Sharpe mass function have been pointed out in
\cite{Grumiller:1999rz}.} Thus, background independence holds only
in the bulk but fails to hold in the asymptotic region; we regard
this actually as an advantage for describing scattering processes
because there is no ``background independent asymptotic
observer''. What one ends up with is a generating functional for
Green functions depending solely on the matter field $\phi$, the
corresponding source $\sigma$ and on the integration constants
mentioned in the previous paragraph
\cite{Grumiller:2002dm}:\footnote{$(\mathcal{D}\phi)$ denotes path
integration with proper measure. In the context of virtual BHs and
tree level amplitudes questions regarding the measure and source
terms for geometry are mostly irrelevant. Therefore, the
generating functional for Green functions simplifies considerably
as compared to the exact case \cite{Grumiller:2002nm}.}
\begin{eqnarray}
W(\si)&=&\int (\mathcal{D}\phi) \exp\Bigl[\I\int \extd^2x
(\mathcal{L}^{\rm eff}+\si\phi)\Bigr]\,, \nonumber\\
\mathcal{L}^{\rm eff}&=&F(\hat{\Phi})\partial_0\phi\partial_1\phi
- \tilde{g}w^\prime(\hat{\Phi})\,.
\label{eq:4.53a}
\end{eqnarray}
The constant $\tilde{g}$ is an effective coupling which turns out
to be inessential and may be absorbed by a redefinition of the
unit of length. For minimal coupling ($F=\rm const.$) $w^\prime$
is the only source for matter vertices. It is a non-polynomial
function, in general. Moreover, the quantity $\hat{\Phi}$, which
is the quantum version of the dilaton $\Phi$, depends not only on
integration constants but also non-locally on matter; to be more
precise, it depends non-locally on $(\partial_0\phi)^2$ due to the
appearance of the operator $\partial_0^{-1}$. Thus, in general
\textit{the effective Lagrangian density \eqref{eq:4.53a} is
non-local and non-polynomial in the matter field.}

\begin{figure}
\epsfig{file=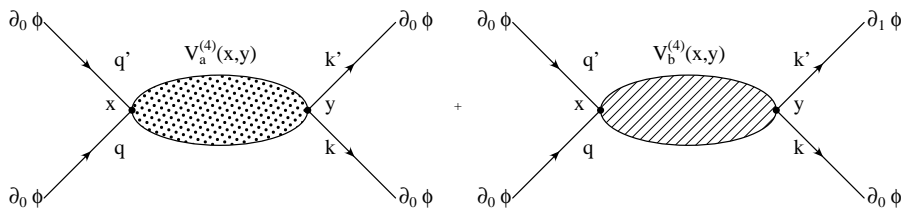,width=0.94\linewidth} \vskip-4mm
\caption{The total $V^{(4)}$-vertex (with outer legs).}
\label{fig2}
\end{figure}

The classical vertices are generated by the conformally invariant
combination of potentials defined in \eqref{eq:entropy14}.
Evidently, constant and linear contributions to the function $w$
are of no relevance for vertices. Perturbation theory yields to
lowest order the two Feynman diagrams depicted in figure
\ref{fig2}. The total $V^{(4)}$-vertex (with outer legs) contains
a symmetric contribution $V^{(4)}_a$ and (for non-minimal
coupling) a non-symmetric one $V^{(4)}_b$. The shaded blobs depict
the intermediate interactions with virtual BHs. With the
abbreviations
$$
\phi_0:=\sfrac12\,(\partial_0\phi)^2\,,\qquad
\phi_1:=\sfrac12\,(\partial_0\phi)(\partial_1\phi)\,,
$$
the Feynman rules for the lowest order non-local vertices read
\cite{Grumiller:2002dm}:\footnote{Eq.~\eqref{vbh:sym} corrects a
misprint in \cite{Grumiller:2002dm} which is irrelevant if
$M_{\mathrm{ADM}}=0$ and therefore was unnoticed until recently
Luzi Bergamin spotted it. I am grateful to him for this
observation. Regarding the label ``ADM'' I wish to express the
caveat that the quantity appearing in \eqref{vbh:sym} need not be
the ADM mass, but for those cases where this notion makes sense it
is related to it in a simple manner.}
$$
V_{a}^{(4)}=-2\int_x\int_y
\phi_0(x)\phi_0(y)\theta(x^0-y^0)\de(x^1-y^1)F(x^0)
F(y^0)\times\hskip20mm$$
\begin{eqnarray}
&&\times\Bigg[4\left(w(x^0)-w(y^0)\right)-2(x^0-y^0)\Big(w'(x^0)+w'(y^0)+\nonumber\\
&&\qquad+\frac{F'(x^0)}{F(x^0)}\left(w(x^0)+
M_{\mathrm{ADM}}\right)+\frac{F'(y^0)}{F(y^0)}\left(w(y^0)+M_{\mathrm{ADM}}\right)\Big)\Bigg]\,,
\label{vbh:sym}
\end{eqnarray}
and
\begin{equation}
V_{b}^{(4)}=-4\int_x\int_y
\phi_0(x)\phi_1(y)\de(x^1-y^1)F(x^0)F'(y^0)
\left|x^0-y^0\right|\,. \label{vbh:asy}
\end{equation}
For the simpler case of minimal coupling, $F=\rm const.$, the
non-symmetric vertex $V_b^{(4)}$ vanishes and the symmetric one
for the ESBH simplifies to
\begin{equation}
\label{eq:entropy101}
\begin{array}{rcl}
V^{(a)}_{\mathrm{ESBH}}&=&\disty
-2F^2\int_x\int_y
\phi_0(x)\phi_0(y)\theta(x^0-y^0)\de(x^1-y^1)\times\\[9pt]
&&\disty\times\left[4\bigl(w(x^0)-w(y^0)\bigr)-
2(x^0-y^0)\biggl(\frac{w(x^0)\ga(x^0)}{\ga^2(x^0)+1}+
\frac{w(y^0)\ga(y^0)}{\ga^2(y^0)+1}\biggr)\right],
\end{array}
\end{equation}
with $w$ as defined in Eqs.~\eqref{eq:entropy18} and \eqref{eq:entropy19}.
\end{appendix}

The $S$-matrix element with ingoing modes $q$, $q'$ and outgoing
ones $k$, $k'$,
\begin{equation}
T(q,q';k,k)=\frac12\left<0\left|a^-_ka^-_{k'}\left(V^{(4)}_a+
V^{(4)}_b\right)a^+_qa^+_{q'}\right|0\right>,
\label{vbh:smatrix}
\end{equation}
depends not exclusively on the vertices (\ref{vbh:sym}),
(\ref{vbh:asy}), but also on the asymptotic states of the scalar
field with corresponding creation/anihilation operators $a^\pm$
obeying canonical commutation relations of the form
$\left[a^-_k,a^+_{k'}\right]\propto \de(k-k')$. Asymptotically the
Klein--Gordon equation with $K:=M_{\mathrm{ADM}}+w(x^0)$ reads
\begin{equation}
\partial_0\left(\partial_1+K\partial_0\right)\phi=
-\frac{F'(x^0)}{2F(x^0)}\,\left(\partial_1+2K\partial_0\right)\phi\,,
\label{vbh:matter}
\end{equation}
If (\ref{vbh:matter}) can be solved exactly and the set of
solutions is complete and normalizable (in an appropriate sense) a
Fock space for incoming and outgoing scalars can be constructed in
the usual way. Since the integration constant $M_{\mathrm{ADM}}$
enters (\ref{vbh:matter}) the asymptotic states depend on the
choice of boundary conditions. Physically, the scalar particles
are scattered on their own gravitational self-energy (the
``Coulomb gravitons'' have been integrated out).

It appears to be straightforward to solve \eqref{vbh:matter} for
minimal coupling and to calculate the $S$-matrix
\eqref{vbh:smatrix} with \eqref{eq:entropy101}. Such a calculation
is of interest (but will not be performed here) because for the
Witten BH \cite{Mandal:1991tz,Callan:1992rs}, which arises as
limit of the ESBH, the function $w$ is linear and therefore
exhibits scattering triviality \cite{Grumiller:2002dm}. Obviously,
the $\al^\prime$ corrections implicit in the ESBH are able to
transcend this triviality result.

\end{document}